# Competition between the Pseudogap and Superconducting States of Bi$_2$Sr$_2$Ca$_{0.92}$Y$_{0.08}$Cu$_2$O$_{8+\delta}$ Single Crystals Revealed by Ultrafast Broadband Optical Reflectivity

G. Coslovich,[1,*] C. Giannetti,[2,3] F. Cilento,[1,4] S. Dal Conte,[2,3] T. Abebaw,[1,†] D. Bossini,[3,‡] G. Ferrini,[2,3] H. Eisaki,[5] M. Greven,[6] A. Damascelli,[7,8] and F. Parmigiani[1,4]

[1]*Department of Physics, Università degli Studi di Trieste, Trieste I-34127, Italy*
[2]*I-LAMP (Interdisciplinary Laboratories for Advanced Materials Physics), Università Cattolica del Sacro Cuore, Brescia I-25121, Italy.*
[3]*Department of Physics, Università Cattolica del Sacro Cuore, Brescia I-25121, Italy*
[4]*Sincrotrone Trieste S.C.p.A., Basovizza I-34012, Italy*
[5]*Nanoelectronics Research Institute, National Institute of Advanced Industrial Science and Technology, Tsukuba, Ibaraki 305-8568, Japan*
[6]*School of Physics and Astronomy, University of Minnesota, Minneapolis, Minnesota 55455, USA*
[7]*Department of Physics & Astronomy, University of British Columbia, Vancouver, British Columbia V6T 1Z1, Canada*
[8]*Quantum Matter Institute, University of British Columbia, Vancouver, British Columbia V6T 1Z4, Canada*
(Dated: January 31, 2013)

Ultrafast broadband transient reflectivity experiments are performed to study the interplay between the non-equilibrium dynamics of the pseudogap and the superconducting phases in Bi$_2$Sr$_2$Ca$_{0.92}$Y$_{0.08}$Cu$_2$O$_{8+\delta}$. Once superconductivity is established the relaxation of the pseudogap proceeds $\sim$ 2 times faster than in the normal state, and the corresponding transient reflectivity variation changes sign after $\sim$ 0.5 ps. The results can be described by a set of coupled differential equations for the pseudogap and for the superconducting order parameter. The sign and strength of the coupling term suggest a remarkably weak competition between the two phases, allowing their coexistence.

In cuprates [1], pnictides [2], manganites [3] and heavy-fermion compounds [4] low-energy gaps follow the formation of multiple ordered phases, generally interacting with each other. Their mutual coupling, i.e., the mixed terms in a multi-component Ginzburg-Landau (GL) free energy expansion, plays a key role in determining the unconventional electronic properties of these materials [5–8]. A particularly interesting case is given by the pseudogap (PG) and superconducting (SC) phases in cuprates, where the debate between a competing order (positive free energy coupling term) [9, 10] or a cooperative order scenario (negative free energy coupling term)[11, 12] is still unresolved.

So far this problem has been approached by measuring the excitation spectra and low-energy gaps of each phase in equilibrium conditions throughout the phase diagram. Angle-resolved photoemission spectroscopy and scanning tunneling spectroscopy have successfully distinguished almost degenerate low-energy gaps in cuprates [13, 14]. However, at equilibrium the coupling between coexisting phases can hardly be observed for the free energy is kept at minimum and the different contributions cannot be disentangled. Conversely, ultrafast time-resolved spectroscopy can address this fundamental issue by investigating the non-equilibrium dynamics of the interacting phases after a sudden non-thermal photoexcitation [15, 16]. Above all, the possibility to create a transient non-thermal phase, where only one of the order parameters is selectively quenched, is critical for directly measuring the coupling between order parameters. In particular, in the limit of the *local equilibrium approximation* [17], and for excitation densities well below saturation [18], the non-equilibrium dynamics of the order parameter and its corresponding excitations coincide [19]. Thus the dynamics of the order parameters can be probed in real-time by measuring the reflectivity variation, $\Delta R/R$. Furthermore, a broad spectral range of probing photon energies is needed to disentangle the dynamics of different interacting order parameters [16].

In this Letter, ultrafast broadband transient reflectivity experiments allow to identify and observe the dynamics of PG and SC phases in underdoped Bi$_2$Sr$_2$Ca$_{0.92}$Y$_{0.08}$Cu$_2$O$_{8+\delta}$ single crystals (T$_C$ = 85 K). Singular Value Decomposition (SVD) is applied to disentangle their spectro-temporal fingerprints in the out-of-equilibrium reflectivity data. At low pump fluence (10 $\mu$J/cm$^2$) the recovery dynamics of the PG proceeds $\sim$ 2 times faster once the SC order parameter is established, and eventually the corresponding transient reflectivity variation changes sign after $\sim$ 0.5 ps. Furthermore, the amplitude of the PG perturbation decreases by a factor 2 below T$_C$. These results prove the existence of a coupled dynamics for the PG and for the SC order parameter, which can be described by a set of differential equations with a coupling term connecting their relaxation dynamics. Under the assumption that the PG phase has a broken symmetry, the time-dependent GL approach provides a framework for interpreting the coupling term as a positive interaction between order parameters in the GL expansion of the free energy. Quantitive estimations suggest that the interaction is weak enough for allowing the coexistence of the two competing phases. This experiment represents a benchmark for the recent developments of theoretical non-equilibrium methods beyond quasi-equilibrium approximations, where the interplay between superconductivity and pseudogap might arise at the quantum level [20].

The dynamics of the non-equilibrium optical response is probed by combining the supercontinuum light generation and detection in the 1.1-2 eV spectral region using a photonic crystal fiber (set-up described in [21]) and an Optical Parametric Amplifier (OPA) with output in the 0.5-1.1 eV range [22]. The experiments are performed on high quality underdoped $Bi_2Sr_2Ca_{0.92}Y_{0.08}Cu_2O_{8+\delta}$ single crystals ($T_c$ = 85 K). The sample growth and annealing methods are the same as previously reported [23, 24].

Figs. 1a and 1b show the transient perturbations of the reflectivity as a function of both time delay and probe photon energy as measured for the PG and SC phases respectively. The experiment is performed at low fluence (10 $\mu J/cm^2$), where the $\Delta R/R$ signal is roughly proportional to the perturbation of the order parameters [17–19]. The 2D matrices, $\Delta R/R(E,t)$, have been decomposed through SVD, i.e., by calculating the corresponding $l$-rank matrix, $\sum_{k=1}^{l} \Delta\phi_k(E)\Delta\psi_k(t)$, that best reproduces the experimental data, where $\Delta\psi_k(t)$ is the temporal eigenfunction normalized to its maximum and $\Delta\phi_k(E)$ is the spectral eigenfunction, containing information about the absolute magnitude of the peak signal at each energy E [22]. This method, which has been widely applied in several research areas [25, 26], yields both energy- and time-domain information, and allows the identification of a minimal number of spectro-temporal components that reproduce a set of data [26]. These components are sorted considering their relative weight, with the first component generally accounting for ∼ 80% of the signal and the second for ∼ 10 %. Higher order components generally represent the experimental noise. In fact, the first component alone gives a fair and almost noise-free representation of the experimental data (see insets of Fig. 1c and 1d).

The spectral eigenfunction of the first component obtained in the PG phase (T = 100 K) is peaked at about 1.2 eV and becomes negative at photon energies higher than 1.4 eV (Fig. 1c). The second component obtained in the PG phase is the remnant normal state signal observed at room temperature [27] and is not relevant in this work. Below $T_C$ the energy-time response changes dramatically with the appearance of a new first component, exhibiting a large positive plateau above 1.2 eV and a sign change below 1.1 eV (Fig. 1d, T = 20 K). This first component originates from photoexcitations across

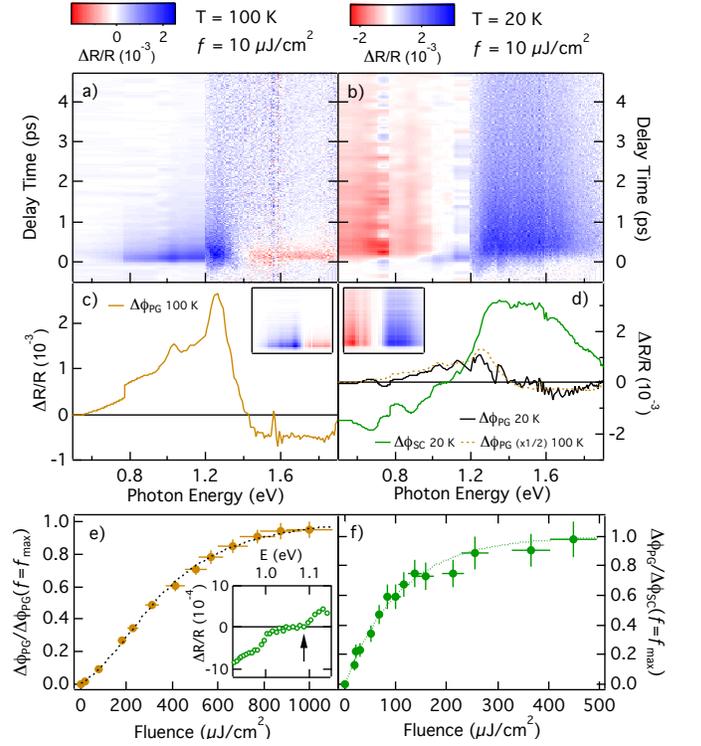

FIG. 1. Time-energy matrices of reflectivity variations at 100 K (a) and 20 K (b) at low fluence (10 $\mu J/cm^2$) on underdoped $Bi_2Sr_2Ca_{0.92}Y_{0.08}Cu_2O_{8+\delta}$. The data are shown in false color scale. c), d) Spectral traces of the SVD components at 100 K and 20 K respectively. The insets show the spectra-temporal matrix of the component with the same scale used in a) and b). e), f) Fluence dependence of the PG and SC components respectively, measured at fixed probe energies (see text). Each experimental point is normalized to the high fluence limit. The dotted lines are fits with a non-linear function [22] exhibiting exponential suppression at low-fluence (e) and saturation at high fluence (e and f) of the form $\propto 1 - e^{-f/f_{sat}}$, with $f_{sat}^{PG} \sim 400$ $\mu J/cm^2$ and $f_{sat}^{SC} \sim 100$ $\mu J/cm^2$. The SC component around 1.1 eV is shown in the inset of (e).

the SC gap causing the transfer of spectral weight in the interband spectral region [24].

In the SC phase the second component of the SVD becomes significant. Remarkably this component exhibits the same spectral shape of the PG signal at 100 K, but with half the amplitude at the same pump fluence (Fig. 1d). Further details can be obtained by considering the probe photon energy where the SC component has a node (1.08 eV, see inset of Fig. 1e), and hence the $\Delta R/R$ signal is dominated by the PG signal. The fluence dependence at this probe photon energy (Fig. 1e) exhibits a saturation threshold ($f_{sat} \sim 400$ $\mu J/cm^2$) similar to the one measured at 100 K, and clearly different from the saturation fluence observed at 0.5 eV (∼ 100 $\mu J/cm^2$, see Fig. 1f), where instead the SC component is maximal.

These results are consistent with a scenario where the PG and SC phases coexist below $T_C$ and compete in the absorption of the pump photon energy. In previous time-resolved single-color experiments at 800 nm (1.5 eV) the residual PG phase component below $T_C$ was hidden by the large SC phase response [18]. These results, while confirming previous experiments reporting the coexistence of two dynamics below $T_C$ [16], allow for the first time the unambiguous and precise measurement of both $\psi_k(t)$ dynamics by recognizing their spectro-temporal fingerprints with no *a priori* assumptions.

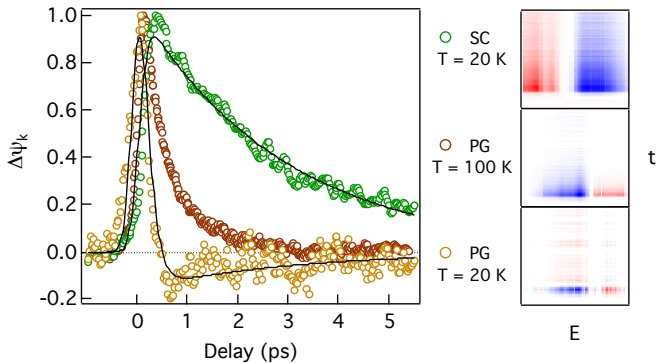

FIG. 2. Temporal traces, $\Delta\psi_k(t)$, of the first two SVD components obtained at T = 20 K, (data of Fig. 1) and the fit obtained by the set of coupled differential equations, Eq. 1. The PG component at 100 K is shown for comparison. As a reference, the insets on the right side show the spectro-temporal matrices of each components with the same scale used in Fig. 1a and 1b.

The results concerning the non-equilibrium dynamics of PG and SC components are shown in Fig. 2. The PG recovery dynamics drastically changes below $T_C$ exhibiting a $\sim 2$ times faster initial decay time and a change of sign after $\sim 500$ fs. Because of this sign change, such dynamics cannot represent a simple quasiparticle density relaxation process. The dynamics of both components can be reproduced by a set of coupled differential equations of the form:

$$\frac{d}{dt}\begin{pmatrix}\Delta\psi_{SC}\\ \Delta\psi_{PG}\end{pmatrix} = \begin{pmatrix}I_{SC}(t)\\ I_{PG}(t)\end{pmatrix} - \begin{pmatrix}\tau_{11}^{-1} & \tau_{12}^{-1}\\ \tau_{21}^{-1} & \tau_{22}^{-1}\end{pmatrix}\begin{pmatrix}\Delta\psi_{SC}\\ \Delta\psi_{PG}\end{pmatrix} \quad (1)$$

where $I_{SC}(t)$ and $I_{PG}(t)$ are the external perturbations induced by the pump pulse, and $\tau_{11}$, $\tau_{22}$, $\tau_{12}$, $\tau_{21}$, represent the diagonal and mixed relaxation terms for the SC (1) and the PG (2) components. The amplitude of the initial perturbations $\Delta\psi_k(t=0)$ can be fixed by considering that the condition of $\Delta\psi_k(t=0)=1$ is reached at the saturation in the high fluence limit (Fig. 1e and Fig.1f), i.e., when the photo-induced vaporization of the ordered phase occurs [18]. The non-equilibrium dynamics of both SC and PG can be simultaneously reproduced by Eq. 1 (Fig. 2). The $\tau_{22}$ value obtained at 20 K is substantially lower than $\tau_{22}$ at 100 K data (decreasing from about $\sim 0.5$ ps to $\sim 0.2$ ps) and a non-zero mixed term, $\tau_{21} \sim 17$ ps, is found. On the other hand $\tau_{12}$ has a negligible effect on the dynamics since the PG component is perturbed one order of magnitude less than the SC component.

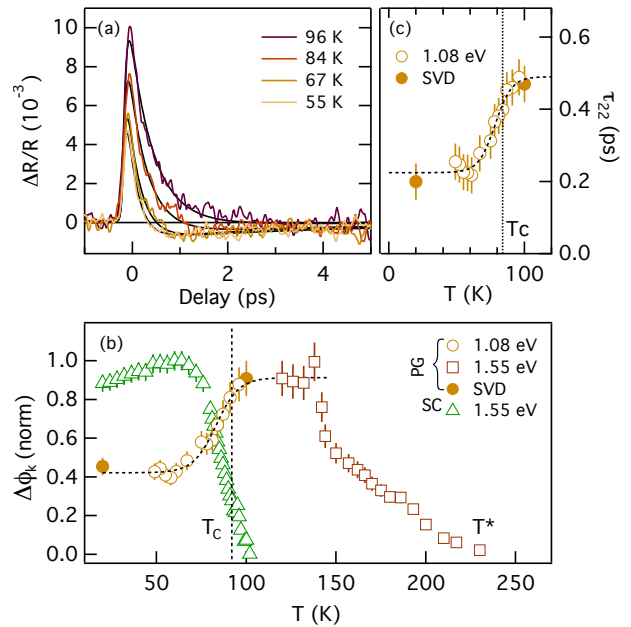

FIG. 3. a) Transient reflectivity variation of the PG component measured at 1.08 eV photon energy and pump fluence of $\sim 50$ $\mu J/cm^2$ as a function of temperature. Black solid lines are fits to Eq. (1). b) Temperature dependence of SC and PG components normalized to the maximal value in the 20-250 K temperature range. c) Temperature dependence of the initial decay time of the PG component. Empty circles, squares and triangles are data taken at a single photon energy, full circles are obtained through SVD analysis of the full spectrum. To improve the signal-to-noise ratio in this low-fluence measurements the PG component is measured at 1.08 eV below $T_C$ and at 1.55 eV above. For the same reason the SC component is measured at 1.55 eV instead of 0.5 eV, as the PG signal remains negligible also at this photon energy in the low-fluence limit. Dashed lines are guides to the eye.

The detailed temperature dependence of the PG component dynamics can be measured at 1.08 eV (Fig. 3), where the highest contrast is obtained over the SC component, as previously discussed. The absence of a long decay signal (for $t > 5$ ps) further confirms that the SC component remains negligible up to $T_C$. The initial relative perturbation of both PG and SC components is shown in Fig.3b as a function of temperature. The trend of the perturbation amplitude of the SC phase is related to the temperature dependence of the equilibrium SC order parameter [18]. Instead the suppression of the PG is inversely proportional to the SC order parameter, with a sharp transition at $T_C$. A similar transition can be observed in the initial relaxation timescale, which becomes $\sim 2$ times faster just below $T_C$ (Fig. 3c) and below the critical fluence, $f_{sat}^{SC}$ [22], i.e., when the superconducting state is non-thermally quenched [18]. Since the non-thermal quenching of the superconducting phase is driven on a timescale faster than the lattice heating, this result demonstrates that the observed changes in the PG dynamics are a genuine result of the interplay between the PG and SC phases without any role played by the lattice temperature.

In summary the main changes to the PG component dynamics that sharply occur once the SC phase sets in are: i) the perturbation decreases by about a factor of two at constant pump fluence (10 $\mu$J/cm$^2$), ii) the initial relaxation dynamics proceed about 2 times faster, iii) a slightly negative variation is observed after about 0.5 ps.

These results can be consistently interpreted within a two-component time-dependent GL model, under the assumption that the pseudogap has a broken symmetry [5, 28, 29]. In facts an increasing number of experiments suggest the presence of an order parameter in the pseudogap phase, i.e., below T* [30, 31]. Remarkably the onset temperature of the PG component measured in this work (Fig. 3b) exactly coincides with the T* reported on similar samples [31]. The nature of this order parameter is still under debate, and the discussion will be kept at the most general level, considering the only assumption that $\psi_{SC}$ and $\psi_{PG}$ are two complex order parameters that break different symmetries. In this case the lowest-order symmetry-allowed GL functional is [8, 29]:

$$F = \alpha_{SC}|\psi_{SC}|^2 + \beta_{SC}|\psi_{SC}|^4 + \alpha_{PG}|\psi_{PG}|^2 + \\ + \beta_{PG}|\psi_{PG}|^4 + W|\psi_{SC}|^2|\psi_{PG}|^2 \quad (2)$$

where the GL expansion is up to quartic order, with $\alpha_k$ and $\beta_k$ as expansion coefficients, and the interaction term couples the squares of the order parameters, with sign and strength determined by $W$. The kinetic equation for each order parameter can be obtained through the relation $d\psi/dt = -\gamma \partial F/\partial \psi^*$, where $\gamma$ is a constant kinetic coefficient [32]. Following Refs. 33 this expression can be linearized for small perturbations of the PG and SC order parameter amplitudes, yielding a set of differential equations of the same form as Eq. 1 which reproduce the data reported in this Letter [22].

This simple model is not comprehensive [33, 34], but it captures the main physics of the coupling between the two relaxation dynamics. The mixed term contains some crucial information, as its sign directly derives from $W$. A positive mixed term implies a repulsive coupling $W$, and in turn a competition between order parameters. Fig.4 shows the non-equilibrium PG phase dynamics predicted by the time-dependent GL model for three different couplings $W$. The results reported here clearly correspond to the repulsive, $W > 0$, case.

To obtain a quantitative estimate of the strength of the coupling it is possible to calculate the dimensionless ratio

$$\frac{\tau_{21}}{\tau_{22}} = \frac{W}{\sqrt{2\beta_{SC}\beta_{PG}}} \frac{\sqrt{\alpha_{SC}\alpha_{PG}} - \mathcal{O}(W)}{-\alpha_{SC} - \mathcal{O}(W)} \quad (3)$$

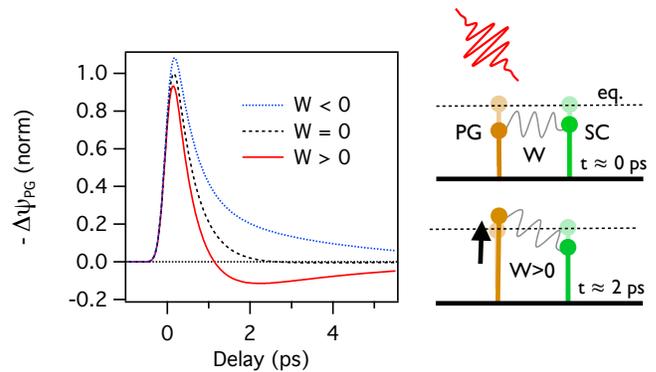

FIG. 4. Dynamics of the PG order parameter obtained from the time-dependent GL model (see text) for the case of three different couplings, repulsive ($W > 0$), attractive ($W < 0$) and no interaction ($W = 0$). The repulsive ($W < 0$) case better reproduces the experimental data (Fig.2 and 3). A schematic representation of the order parameter dynamics after photoexcitation is shown on the right side. The height of the brown (green) symbols indicates the amplitude of the PG (SC) order parameters. Due to the different relaxation timescales, after $\sim 2$ ps (lower panel) only the PG order parameter is close to the equilibrium value and exhibits a small enhancement due to the repulsive interaction with the SC order parameter.

where the terms of higher order, $\mathcal{O}(W)$, can be neglected for small $W$. By considering the experimental values of $\tau_{22}$ and $\tau_{21}$ and by using the parametrization reported in [9], it is possible to estimate the ratio $W/\sqrt{\beta_{SC}\beta_{PG}} \sim 10^{-2}$. A ratio smaller than 1 indicates that phase competition is weak enough to allow the coexistence of PG and SC phases [8], therefore rationalizing the apparent dichotomy of coexistence and competition reported in the literature [10, 12, 15].

In conclusion, ultrafast broadband transient reflectivity experiments allow to identify and measure the dynamics of PG and SC phases in underdoped Bi$_2$Sr$_2$Ca$_{0.92}$Y$_{0.08}$Cu$_2$O$_{8+\delta}$ single crystals (T$_c$ = 85 K), by disentangling their spectro-temporal signatures in the out-of-equilibrium reflectivity data. The results prove the existence of a coupled dynamics for the PG and for the SC order parameter. The data, interpreted within a time-dependent GL approach under the assumption that the PG phase has a broken symmetry, suggest a weak positive interaction between the order parameters, leading to the coexistence of the two competing phases. The ability to quantitatively estimate these coupling parameters paves the way for further applications of ultrafast broadband reflectivity to a variety of complex systems characterized by interacting, yet coexisting, phases [2–4, 6, 7].

We acknowledge discussions and suggestions from M. Schirò, M. Capone and A. Avella. F.C., G.C., and F.P. acknowledge the support of the Italian Ministry of University and Research under Grant Nos. FIRBR-

BAP045JF2 and FIRB-RBAP06AWK3. The research activities of C.G., S.D.C. and F.P. have received funding from the European Union, Seventh Framework Programme (FP7 2007-2013), under grant agreement nr. 280555. The crystal growth work was performed in M.G.s prior laboratory at Stanford University, Stanford, CA 94305, USA, and supported by DOE under Contract No. DE-AC03-76SF00515. The work at UBC was supported by the Alfred P. Sloan Foundation (A.D.), the CRC, Killam, and NSERC's Steacie Fellowship Programs (A.D.), NSERC, CFI, CIFAR Quantum Materials, and BCSI.

# Supplementary Materials for
## "Competition between the pseudogap and superconducting states of $Bi_2Sr_2Ca_{0.92}Y_{0.08}Cu_2O_{8+\delta}$ single crystals revealed by ultrafast broadband optical reflectivity"

G. Coslovich,[1] C. Giannetti,[2,3] F. Cilento,[1,4] S. Dal Conte,[2,3] T. Abebaw,[1] D. Bossini,[3] G. Ferrini,[2,3] H. Eisaki,[5] M. Greven,[6] A. Damascelli,[7,8] and F. Parmigiani[1,4]

[1] Department of Physics, Università degli Studi di Trieste, Trieste I-34127, Italy
[2] I-LAMP (Interdisciplinary Laboratories for Advanced Materials Physics), Università Cattolica del Sacro Cuore, Brescia I-25121, Italy.
[3] Department of Physics, Università Cattolica del Sacro Cuore, Brescia I-25121, Italy
[4] Sincrotrone Trieste S.C.p.A., Basovizza I-34012, Italy
[5] Nanoelectronics Research Institute, National Institute of Advanced Industrial Science and Technology, Tsukuba, Ibaraki 305-8568, Japan
[6] School of Physics and Astronomy, University of Minnesota, Minneapolis, Minnesota 55455, USA
[7] Department of Physics & Astronomy, University of British Columbia, Vancouver, British Columbia V6T1Z1, Canada
[8] Quantum Matter Institute, University of British Columbia, Vancouver, British Columbia V6T1Z4, Canada

## 1. Experimental Methods

The setup for the near-infrared pump-probe experiment utilizes a 50-fs Ti:Sapphire Regenerative Amplified Laser (RegA) operating at a repetition rate as high as 250 kHz, and an Optical Parametric Amplifier (OPA) generating tunable ultrashort pulses in the 0.5-1 eV spectral range. The pump beam (800 nm from the RegA) delay time is controlled with two high-precision motorized translators, a 30 cm long delay stage (PI) for slow movements and a compact delay stage (APE) for continuous movements at frequencies of about 1-2 Hz. The precision of both delay stages is of the order of 1 μm.

A motorized variable attenuator precisely controls via software the intensity of the pump pulse, while a mechanical high-speed chopper allows the high-frequency pump modulation for lock-in detection. The main advantage of high-speed modulation is to minimize the *1/f* shot noise in the detection. In the experiments reported here the modulation frequency was set to 13 kHz. The pump beam is then focused on the sample with a plano-convex lens with focus of 300 mm.

The tunable wavelength probe beam from the OPA is properly attenuated to avoid self-induced effects on the sample and polarized in the ab-plane of the sample. The pulse is then focused on the sample with a silver spherical mirror with effective focal length of 200 mm. The use of the spherical mirror instead of a plano-convex lens minimizes the beam position variations at different wavelengths. A stage for second harmonic generation in BBO can be used to double the probe frequency and reach the wavelength range of 700-550 nm. Measurements in this spectral range allow the exact connection with the Supercontinuum pump-probe experiments (1-2 eV range).

The size of the pump and probe beams on the sample are measured by imaging the spots in the focal plane of a CCD camera. The pump spot is generally about 90-100 μm, while the probe spot varies between 50-80 μm, with an uncertainty of ≈ 5% on each measurement. This causes an uncertainty of ≈ 10% on the determination of the pump fluence impinging the sample.

The sample is placed in an open-cycle cryostat, which cool down the sample to a minimum temperature of 10 K. The detection is based on a single photodiode (InGaAs or Si depending on the frequency range) and subsequent lock-in amplification of the signal modulated at the chopper frequency (13 kHz). A Solar M266-iv imaging monochormator is also used to increase the spectral resolution of the measurements.

Supercontinuum measurements in the 1-2 eV spectral range have been performed using the setup previously described in Ref. [1]. The repetition rate of the laser source is set at 100 kHz to avoid laser-induced average heating effects and has been increased for low-fluence measurements to improve the signal-to-noise ratio. The temporal resolution of the probe pulses varies with the wavelength, with an average value of 100 fs, and it has been characterized via two-photon absorption on a ZnSe plate and via sum-frequency in a BBO crystal. The absolute reference of time-zero is taken in respect to the very high-fluence data (> 1 mJ/cm$^2$), which exhibit the same fast rise time at all temperatures and wavelengths.

The noise of the ΔR/R(ω,t) signal varies depending on the experimental conditions, such as integration time, chopping frequency, laser repetition rate used and wavelength. Moreover some fluctuations (≈ 5%) of the effective pump fluence, mostly due to spatial overlap optimization, may be present. The overall uncertainty on ΔR/R(ω,t) is thus estimated as a sum of the *rms* noise (measured at negative delay times, *t* < 0) and the 5% uncertainty on pump fluence fluctuations. Error analysis on the fitting procedure allows estimating the uncertainties of the calculated decay times.

## 2. Methods: SVD decomposition

In the spectroscopic time-resolved experiments performed in this Letter the outcome is a time-energy matrix ΔR/R(ω,t). There are several approaches to the data analysis of this matrix. One can consider for example the spectral traces ΔR/R(ω) obtained at fixed delay time *t*, analyze them separately with a differential fitting model and then later reconstruct a time-dependence of each particular parameter used to analyze the spectral data. This method provides very interesting information on the dynamical spectral weight transfer [2], but it is less transparent when disentangling coexisting perturbations of the dielectric function.

An alternate approach would start instead from the dynamical trace $\Delta R/R(t)$ integrated on a narrow frequency range $\Delta\omega$. By fitting the time trace with multiple exponentials one can distinguish different components by their different relaxation rates. This method has been successful in showing the existence of two coexisting components in Bi2212 [3], but becomes very challenging when one of the components exhibits a slow rise time (and this is the case of the superconducting state in Bi2212 [4]). In facts, component separation through this method did not provide reliable information about the dynamics for $t < 1$ ps in unconventional superconductors [5,6].

These issues can be solved by adopting the Singular Value Decomposition (SVD) method for the $\Delta R/R(\omega,t)$ matrix. In this case both the spectral ($\Delta R/R(\omega)$) and temporal ($\Delta R/R(t)$) information are used to properly decompose the experimental data. In the following, we review this method, considering an $MxN$ data matrix, called $X$.

The SVD procedure decomposes the data matrix $X$ into a product of three matrices,

$$X = USV^T$$

where U is an $MxN$ matrix whose columns are called the left singular vectors, $\mathbf{u}_k$, and form an orthonormal basis for the spectral traces $\Delta R/R(\omega)$, S is a diagonal matrix whose elements $s_k$ are called singular values and $V^T$ is an $NxN$ matrix containing the elements of the right singular vectors, $\mathbf{v}_k$, and form an orthonormal basis for the dynamical traces $\Delta R/R(t)$. By convention, the ordering of the singular vectors is determined following high-to-low sorting of singular values, with the highest singular value in the upper left index of the $S$ matrix. For a square, symmetric matrix $X$, singular value decomposition is equivalent to the solution of the eigenvalue problem. Moreover the SVD method gives the same result of the Principal Component Analysis (PCA) when the principal components are extracted from the covariance matrix [7].

The SVD method can be applied to find a lower-rank matrix, which best reproduces $X$,

$$X^{(l)} = \sum_{k=1}^{l} \mathbf{u}_k s_k \mathbf{v}_k$$

where $X^{(l)}$ is the closest rank-$l$ matrix to $X$. The term closest means that $X^{(l)}$ minimizes the sum of the squared differences of the elements of $X$ and $X^{(l)}$, $\sum_{ij}|x_{ij} - x_{ij}^{(l)}|^2$.
SVD has been successfully used in a number of areas such as the analysis of spectroscopic data [8], of the temporal variation of genome-wide expression [9] and of time-resolved macromolecular x-ray experiments [10].
In this work, SVD has been applied, using LAPACK routines, to analyze the time-energy matrices $\Delta R/R(\omega, t)$. The matrices are generally well reproduced by the first two components, with the first one generally containing about the 80 % of the signal of the original matrix.

It should be noted that the SVD method relies on the assumption that the experimental data can be represented as a linear sum of independent components. While this assumption is not necessarily true for all experimental conditions, it is reasonable to expect that at low enough pump fluence the system would respond linearly to photo-excitation. A confirmation of the validity of this assumption is evident from Fig. 1 of this Letter, where the spectral features of the PG component are reproduced almost exactly above and below $T_C$, despite a strong change in the amplitude of the other components.

In conclusion the SVD is a powerful tool to recognize the fundamental spectral responses, $\mathbf{u}_k$ or equivalently $\Delta R_k/R(\omega)$, and the corresponding temporal evolution, $\Delta R k/R(t)$ in the experimental data. Each spectral response $\Delta R_k/R(\omega)$ and temporal trace $\Delta R k/R(t)$ are associated with the perturbation of a distinct order parameter k, which produces a particular kind of perturbation of the dielectric function on a certain timescale.

## 3. Fluence dependence

The fluence dependence of the PG component amplitude (Fig. 1e) is reproduced by the phenomenological formula

$$\Delta\phi_{PG}(f) = A(1 - e^{-f/f_{sat}^{PG}})(B + \frac{1}{1 + e^{(f_{sat}^{SC}-f)/f_0}})$$

where the first term accounts for the saturation at high fluence and the second one (a sigmoid function) reproduces the sub-linear suppression at low fluence. The suppression is due to the competition between the two order parameters in absorbing the pump fluence in this regime. $f_{sat}^{SC}$ is fixed to the saturation fluence of the SC component, while $A$, $B$, $f_{sat}^{PG}$ and $f_0$ are free parameters of the fit. A sigmoid function is also used as a guide to the eye in Fig. 3 and in Fig. S1.

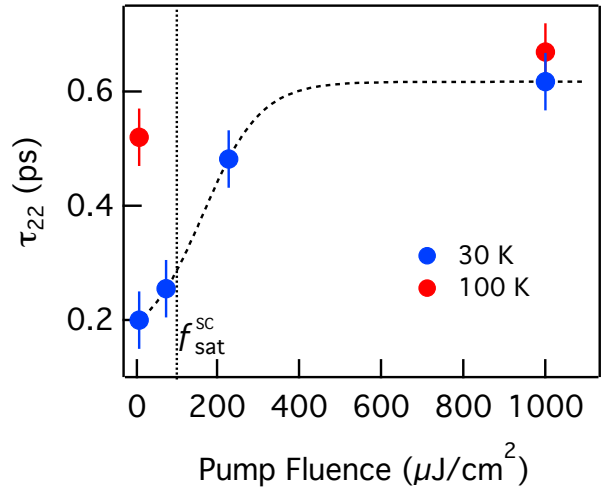

Fig. S1 Fluence dependence of the initial decay time of the PG components obtained from SVD at 30 K (SC phase) and 100 K (PG phase). The dashed line is a guide to the eye.

## 4. Application of Time-dependent Ginzburg-Landau Model

The application of the time-dependent Ginzburg-Landau model developed in this Letter takes into account small variations of two complex scalar order parameters, $\psi_{SC}$ and $\psi_{PG}$, which breaks distinct symmetries. The symmetry-allowed lower order Ginzburg-Landau functional is [11,12]:

$$F = \alpha_{SC}|\psi_{SC}|^2 + \beta_{SC}|\psi_{SC}|^4 + \alpha_{PG}|\psi_{PG}|^2 + \beta_{PG}|\psi_{PG}|^4 + W|\psi_{SC}|^2|\psi_{PG}|^2$$

where the Ginzburg-Landau expansion is up to the quartic order, with $\alpha_k$ and $\beta_k$ as expansion coefficients, and the interaction term couples the squares of the order parameters, with sign and strength determined by W.

The kinetic equation for the pseudogap order parameter becomes

$$\frac{d\psi_{PG}}{dt} = -\lambda \frac{\delta F}{\delta \psi_{PG}^*} = -\lambda\{\alpha_{PG}\psi_{PG} + 2\beta_{PG}|\psi_{PG}|^2\psi_{PG} + W|\psi_{SC}|^2\psi_{PG}\}$$

where $\lambda$ is a constant kinetic coefficient. Considering a perturbation of the order parameters, $\psi_{SC} = \bar{\psi}_{SC} + \delta\psi_{SC}$ and $\psi_{PG} = \bar{\psi}_{PG} + \delta\psi_{PG}$, one can linearize the expression for small perturbations. If only amplitude perturbations of the order parameters (phase perturbations can be neglected within 1 ps) are considered we can obtain a differential equation for the pseudogap order parameter

$$\frac{d\delta\psi_{PG}}{dt} = -4\lambda\left\{(-\alpha_{PG} - W|\bar{\psi}_{SC}|^2)\delta\psi_{PG} + W\bar{\psi}_{PG}\bar{\psi}_{SC}\delta\psi_{SC}\right\}$$

from which we can extract the direct ($\tau_{22}$) and mixed ($\tau_{21}$) terms of equation (1) in the Letter:

$$\tau_{22}^{-1} = 4\lambda\left\{-\alpha_{PG} - W|\bar{\psi}_{SC}|^2\right\}$$

$$\tau_{21}^{-1} = 4W\lambda\left\{\sqrt{\frac{\left(-\alpha_{SC} - W|\bar{\psi}_{PG}|^2\right)\left(-\alpha_{PG} - W|\bar{\psi}_{SC}|^2\right)}{2\beta_{SC}\beta_{PG}}}\right\}$$

To minimize the number of unknown quantities it is useful to evaluate the ratio

$$\frac{\tau_{22}}{\tau_{21}} = \frac{W}{\sqrt{2\beta_{SC}\beta_{PG}}} \frac{\sqrt{\alpha_{SC}\alpha_{PG} - \mathcal{O}(W)}}{-\alpha_{PG} - \mathcal{O}(W)}$$

where $\mathcal{O}(W)$ indicates terms of higher order in W. In facts it is reasonable to expect that W is smaller than $\sqrt{\beta_{SC}\beta_{PG}}$, as in the case of competing orders which allows partial coexistence [11], thus higher orders in W can be neglected. The self-consistency of this hypothesis will be checked *a posteriori*.

In this case the $\tau_{22}/\tau_{21}$ ratio can be approximated with

$$\frac{\tau_{22}}{\tau_{21}} \sim \frac{W}{\sqrt{\beta_{SC}\beta_{PG}}} \sqrt{\frac{\alpha_{SC}}{2\alpha_{PG}}}$$

Following the parameterization proposed by Chakravarty et al. in Ref. [12] the ratio $\alpha_{SC}/\alpha_{PG}$ is about ½ for the underdoped sample under consideration (x=0.128). Thus the $\tau_{22}/\tau_{21}$ ratio gives a direct estimation of the order of magnitude of W, as compared to $\sqrt{\beta_{SC}\beta_{PG}}$. The result of a ratio $\frac{W}{\sqrt{\beta_{SC}\beta_{PG}}} \sim 10^{-2}$ further strengthen the earlier assumption of a small interaction term.